\documentclass[%
 reprint,
 amsmath,amssymb,
 aps,
]{revtex4-2}

\usepackage{graphicx}
\usepackage{dcolumn}
\usepackage{bm}
\usepackage{xcolor}
\usepackage{soul}  
\usepackage{amsmath}
\usepackage{hyperref}
\hypersetup{colorlinks=true, citecolor=blue}

\usepackage{verbatim}

\newcommand{\rev}[1]{{\color{black} #1}}  

\begin{document}
\preprint{APS/123-QED}

\title{A Physical Interpretation of Imaginary Time Delay}

\author{Isabella L. Giovannelli}
 \email{igiovann@umd.edu}
\affiliation{%
Maryland Quantum Materials Center, Department of Physics\\
 University of Maryland, College Park, Maryland 20742, USA
}%

\author{Steven M. Anlage}

\affiliation{%
Maryland Quantum Materials Center, Department of Physics\\
 University of Maryland, College Park, Maryland 20742, USA
}%

\date{\today}

\begin{abstract}
The scattering matrix $S$ linearly relates the vector of incoming waves to outgoing wave excitations, and contains an enormous amount of information about the scattering system and its connections to the scattering channels.  Time delay is one way to extract information from $S$, and the transmission time delay $\tau_T$ is a complex (even for Hermitian systems with unitary scattering matrices) measure of how long a wave excitation lingers before being transmitted.  The real part of $\tau_T$ is a well-studied quantity, but the imaginary part of $\tau_T$ has not been systematically examined experimentally, and theoretical predictions for its behavior have not been tested.  Here we experimentally test the predictions of Asano, \textit{et al}. [Nat. Comm. \textbf{7}, 13488 (2016)] for the imaginary part of transmission time delay in a non-unitary scattering system. We utilize Gaussian time-domain pulses scattering from a 2-port microwave graph supporting a series of well-isolated absorptive modes to show that the carrier frequency of the pulses is changed in the scattering process by an amount in agreement with the imaginary part of the independently determined complex transmission time delay, $\text{Im}[\tau_T]$, from frequency-domain measurements of the sub-unitary $S$ matrix.  Our results also generalize and extend those of Asano, \textit{et al}., establishing a means to predict pulse propagation properties of non-Hermitian systems over a broad range of conditions.


\end{abstract}

\maketitle


\textbf{Introduction.} In linear scattering systems, the scattering matrix $S$ is used to relate incoming waves $|\psi_{in}\rangle$ to outgoing waves $|\psi_{out}\rangle$ where $|\psi_{out}\rangle = S|\psi_{in}\rangle$. The scattering matrix $S$ is a complex function of energy (or frequency) and is a square $M\times M$ matrix where $M$ is the number of channels coupling the system to the outside world. This formulation of scattering as well as its statistical treatment using random matrix theory \cite{verbaarschot1985,sokolov1989,schehr2017,nock2014,mitchell2010,mello1985,fyodorov1997,fyodorov2005,fyodorov2015} can be applied to a wide array of complex systems. A non-exhaustive list includes: microwave and sound scattering experiments \cite{doron1990,richter2001,kuhl2013,hul2012,gradoni2014,dietz2015,kuhl2005}, nuclear and atomic scattering \cite{mitchell2010}, and scattering in quantum many-body systems \cite{bereczuk2021}. The scattering matrix encapsulates a vast amount of information regarding the scattering system \cite{agassi1975,gao2009,weidenmuller1992,mitchell2010}. It can be used to determine how long a wave stays in the scattering system before leaving, which is referred to as time delay.

In the same way that the scattering matrix can be used to describe a broad range of scattering phenomena, time delay is just as widely applicable.  In quantum mechanics, time delay is directly related to the phase evolution of quantum waves \cite{Trabert2021,Zhang2011}. It can also be related to the density of states of open scattering systems \cite{Kuipers2014,Davy2015}. In photonics, time delay can be used to determine group delay in optical fibers and manipulate the shape of wavefronts \cite{carpenter2015,xiong2016,bohm2018,gerardin2016,brandstotter2019}. The time delay operator can also be utilized to optimize light storage within disordered media \cite{durand2019}, and to characterize scattering of narrow-band acoustic pulses \cite{patel2023}. In electromagnetics, time delay can be used to determine group delay in wave guides \cite{fan2005,patel2021,mao2023} and to control the level of energy focused within a microwave enclosure \cite{ambichl2017}. It can also be used to determine the locations of poles and zeros of the scattering matrix in the complex frequency plane \cite{fyodorov1997,fyodorov2017,fyodorov2019,osman2020,chen2021, chen2022}. 
\textbf{Time Delay in Unitary Scattering Systems.} Time delay was first described by Eisenbud \cite{Eisenbud1948} and Wigner \cite{Wigner1955} in the context of elastic nuclear scattering. This concept was later generalized by Smith \cite{Smith1960} to include inelastic scattering and systems with many channels. In the case of classical electromagnetic waves, the setting for the experimental results in this paper, time delay is related to the derivative of the classical wave's scattering phase shift with respect to frequency \cite{huang2022,Davy2015,doron1990}. Written in terms of the frequency dependent scattering matrix, the Wigner-Smith time delay for electromagnetic waves is $\tau_W(\omega) = -\frac{i}{M}\frac{d}{d\omega}\text{ln[det}S(\omega)]$ where $\omega$ is angular frequency.

The statistical properties of time delay in highly-overmoded  unitary scattering systems have been investigated in detail \cite{Lehmann1995,Gopar1996,Misirpashaev1997,Fyodorov1997Cross,fyodorov1997,Fyodorov1998,Tiggelen1999,Brouwer1999,Savin2001,Kottos2003,Mezzadri2013,Texier2013,Novaes2015,Cunden2015,Huang2020}, including its use in quantum transport theory \cite{Texier2016}.  We note that furtive attempts to define a complex generalization of time delay in the context of tunneling \cite{Poll84,Land94} have proven to be of limited physical utility \cite{Win06}.

\begin{figure}[ht!]
    \centering
    \includegraphics[width=0.4\textwidth]{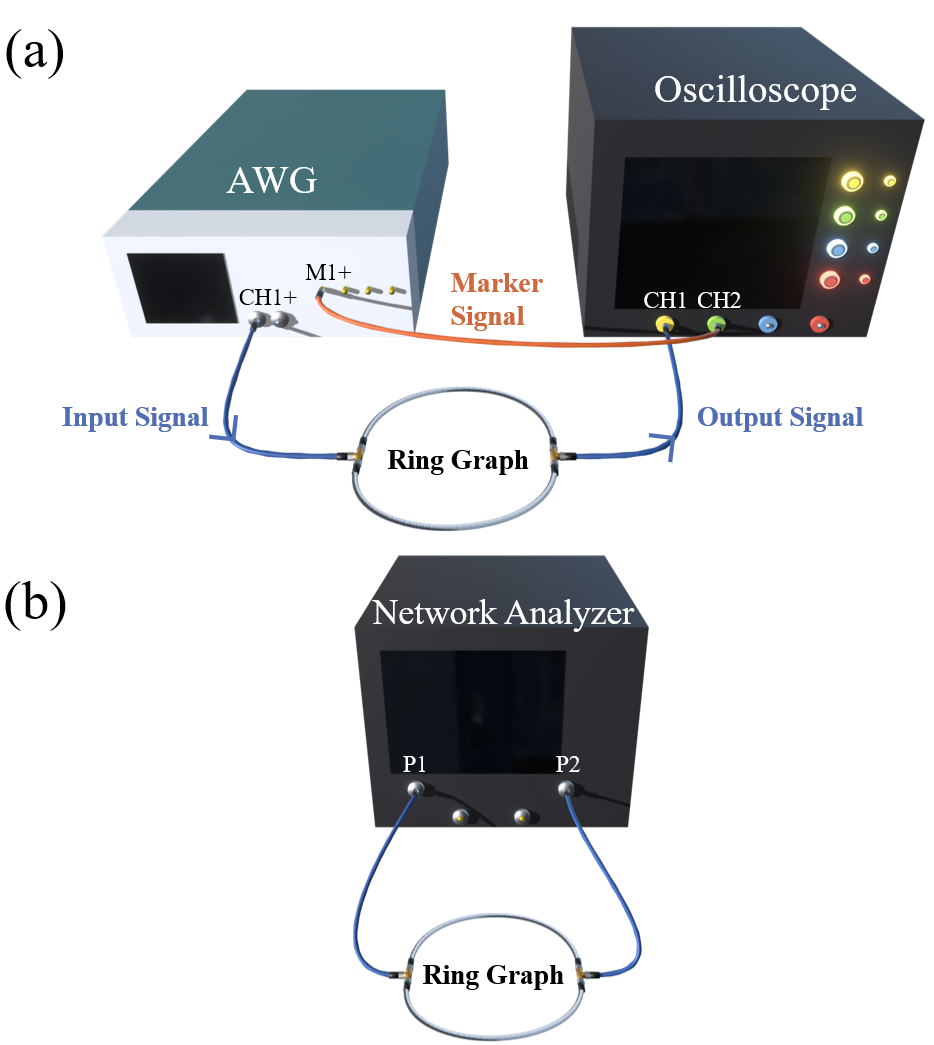}
    \caption{(a) Time-domain experiment setup. (b) Frequency-domain experiment setup.}
    \label{exp_schematic}
\end{figure}
\textbf{Time Delay in Subunitary Scattering Systems.} In this paper, we will focus on the time delay associated with the transmission components of the scattering matrix referred to as transmission time delay ($\tau_T$). This is in contrast to Wigner-Smith time delay which \rev{utilizes the entire scattering matrix}. Since we study a two-port ring graph, the scattering matrix is rank 2: $ S=\begin{pmatrix} R\quad\; T' \\ T\quad R'\end{pmatrix}$ where $T$ and $T'$ are the transmission coefficients and $R$ and $R'$ are the reflection coefficients. Transmission time delay \rev{$\tau_T =  \tau_T(\omega;\alpha)$} is defined generally as \cite{chen2022},
 \begin{align}
     \tau_T &=-i\frac{\partial}{\partial\omega}\text{ln[\text{det}}T(\omega+i\alpha)] = \text{Re}[\tau_T]+i\ \text{Im}[\tau_T] \label{eq:transtime1}
 \end{align}
where $T=|S_{21}|e^{i\phi}$, and $\alpha$ quantifies the uniform loss in the system. Transmission time delay can be analogously defined for $T'$.

The transmission sub-matrix $T$ is sub-unitary, hence $\tau_T$ is complex valued, and its real and an imaginary parts can be either positive or negative. This naturally leads to the question of how to physically interpret this quantity. 
 
Negative real time delay was examined theoretically by Garrett and McCumber \cite{garrett1970} and experimentally demonstrated by Chu and Wong for light pulses interacting with a single isolated absorptive mode \cite{chu1982}. 
Negative time delay occurs when the group velocity ($v_g$) of the pulse surpasses the speed of light $c$, \rev{or becomes negative. This can occur in regions of large anomalous dispersion (e.g. the system is excited near one or more narrow resonances) \cite{boyd2002}. In the $v_g>c$ case the peak of the pulse traveling through an anomalously dispersive medium arrives before an equivalent pulse traveling through vacuum \cite{stenner2003}. In the $v_g<0$ case the peak of the pulse 
leaves the medium before the peak of the incident pulse  enters \cite{gehring2006}. This unintuitive phenomena is the result of inhomogeneous distortion of the Fourier components of the pulse as it travels through the medium, causing shifts in the center of the pulse and its leading edge. The overall pulse character is maintained as long as the frequency bandwidth of the pulse is smaller than the width of the resonance being excited in the medium \cite{garrett1970,boyd2002}. Remarkably, negative real part of complex time delay is also observed in media with gain \cite{Wang00}, as well as loss, and in systems with nonlinear wave mixing \cite{Borto10}.  Recently, in a purely quantum mechanical measurement of single photons traveling through a cloud of resonantly absorbing atoms a negative real part of time delay was observed \cite{angulo2024}, and its value is equal to the group delay (i.e. the real part of Eq.~\ref{eq:transtime1}), suggesting a deep connection between complex time delay and quantum weak measurements.}

Imaginary time delay was first interpreted by Asano \textit{et al}.~\cite{Asano2016} as a center-frequency shift in the pulse rather than a time shift.  
They note that this relationship is similar to that between \rev{frequency shifts and angular Goos-H{\"a}nchen shifts \cite{balcou1997,kogelnik1974,chan1985,merano2009,bliokh2013}}, as well as \rev{frequency shifts and the imaginary part of quantum weak measurement values} \cite{Aharonov88,steinberg1995,aharonov2003,Solli04,brunner2004,brunner2010}.  Asano, \textit{et al.}, make the theoretical connection between imaginary time delay and pulse center frequency shift but do not present corresponding experimental results. In this paper, we extend this work by presenting the corresponding experimental results directly demonstrating the relationship between imaginary time delay and pulse center-frequency shift.

This paper is structured as follows. First, we briefly review the theoretical model describing pulse propagation through dispersive media. We then present the experimental setup, data, and results. These experimental results are directly compared to the predictions made by Asano \textit{et al}. \cite{Asano2016}, and we discuss how our results generalize theirs.

\textbf{Transmission Time Delay and Gaussian Pulse Properties.} To derive the predicted results one can combine methods used in Asano \textit{et al}. \cite{Asano2016} and Cao \textit{et al}. \cite{cao2003}. The calculation details are presented in section III of the Supp. Mat.~\cite{SuppMat}. Here we summarize the highlights. The main assumptions needed are: 1) The frequency bandwidth of the pulse $\tilde{\Delta}_\omega$ is much smaller than the  3-dB linewidth of the resonant mode being studied $\gamma_{3-\text{dB}}$, and 2) \rev{the system is linear and dispersive.}

The predicted shift in transmission time ($D_t$) and center frequency ($D_\omega$) of a transmitted Gaussian pulse is,
\noindent
\begin{minipage}{0.45\linewidth}
  \begin{equation}
    \qquad D_t = \text{Re}[\tau_T]
    \label{eq: Dt}
  \end{equation}
\end{minipage}%
\hspace{0.05\linewidth}
\begin{minipage}{0.45\linewidth}
  \begin{equation}
    D_\omega = -\tilde{\Delta}^2 \, \text{Im}[\tau_T]
    \label{eq: Dw}
  \end{equation}
\end{minipage}

\vspace{0.3cm}
\noindent where \rev{$\tilde{\Delta}=\frac{\tilde{\Delta}_\omega}{2\sqrt{2\text{ln}2}}$ and $\tilde{\Delta}_\omega$ is the full width at half maximum (FWHM) of the pulse Gaussian distribution in the frequency domain. See section V in the Supp. Mat. \cite{SuppMat} for more details.}


\begin{figure}[ht!]
    \includegraphics[width=0.48\textwidth]{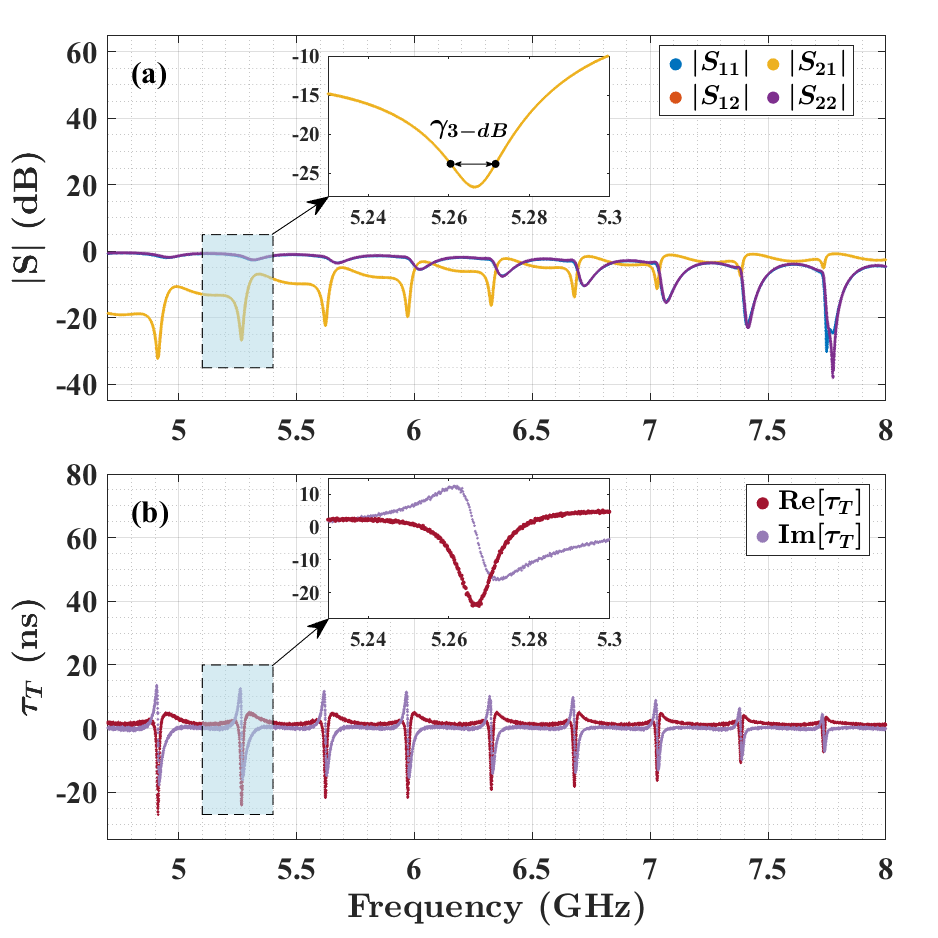}
    \caption{(a) Measured scattering matrix elements for the ring graph depicted in Fig.\ \ref{exp_schematic}(b). The transmission parameters are in orange and yellow (overlapping), the reflection parameters are in blue and purple. The inset is a zoomed-in graph of $S_{21}$ for the indicated boxed region (5.23-5.3 GHz). The 3-dB bandwidth of this resonance is $\gamma_{3-dB}$ = 11.15 MHz. (b) The transmission time delay is calculated using $S_{21}$ data in (a). The real part is plotted in red and the imaginary part of the transmission time delay is plotted in light purple.} 
    \label{SParam_TransTime}
\end{figure}
\textbf{Experiment.} The experiments were performed using a 2-port microwave ring graph as the scattering system. The ring graph is composed of two coaxial cables of different lengths (27.9 and 30.5 cm long) and two T-junctions, and is depicted in both panels of Fig. \ref{exp_schematic}. There are multiple reasons why we found it advantageous to use a ring graph for this experiment. One is that the ring graph has widely spaced and isolated absorptive modes (see Fig. \ref{SParam_TransTime}(a)), allowing for straightforward analysis and interpretation.  Another reason is because the $S$-matrix and complex time delay of the ring graph have already been thoroughly characterized \cite{Waltner13,Waltner14,chen2022}. 
We note in passing that prior work has demonstrated that time delay of short pulses in microwave graphs contains useful information about the structure of the graph \cite{BialPulse21,BialPulse23}.

\textbf{Transmission Time Delay Measurements.} To find the transmission time delay, we used the frequency domain experiment setup depicted in Fig.\ \ref{exp_schematic}(b). Port 1 (P1) of a Keysight N5242A network analyzer (PNA-X) is attached to one end of the ring graph, the other end of the ring graph is attached to port 2 (P2). The PNA-X is calibrated up to the connection points to the ring-graph with a Keysight N4691-60001 Electronic Calibration kit over the 10 MHz to 18 GHz frequency range with a frequency step size of 179.9 kHz. 

Representative frequency domain results are summarized in Fig.~\ref{SParam_TransTime}, where both the measured scattering parameters and the corresponding calculated transmission time delay (using Eq.~\ref{eq:transtime1}) are depicted as a function of frequency. We see in Fig.~\ref{SParam_TransTime}(a) that the modes are widely spaced without any overlap as characterized experimentally in \cite{chen2022}, and assumed theoretically \cite{garrett1970,Asano2016}. 

\begin{figure}[ht!]
    \centering
    \includegraphics[width=0.48\textwidth]{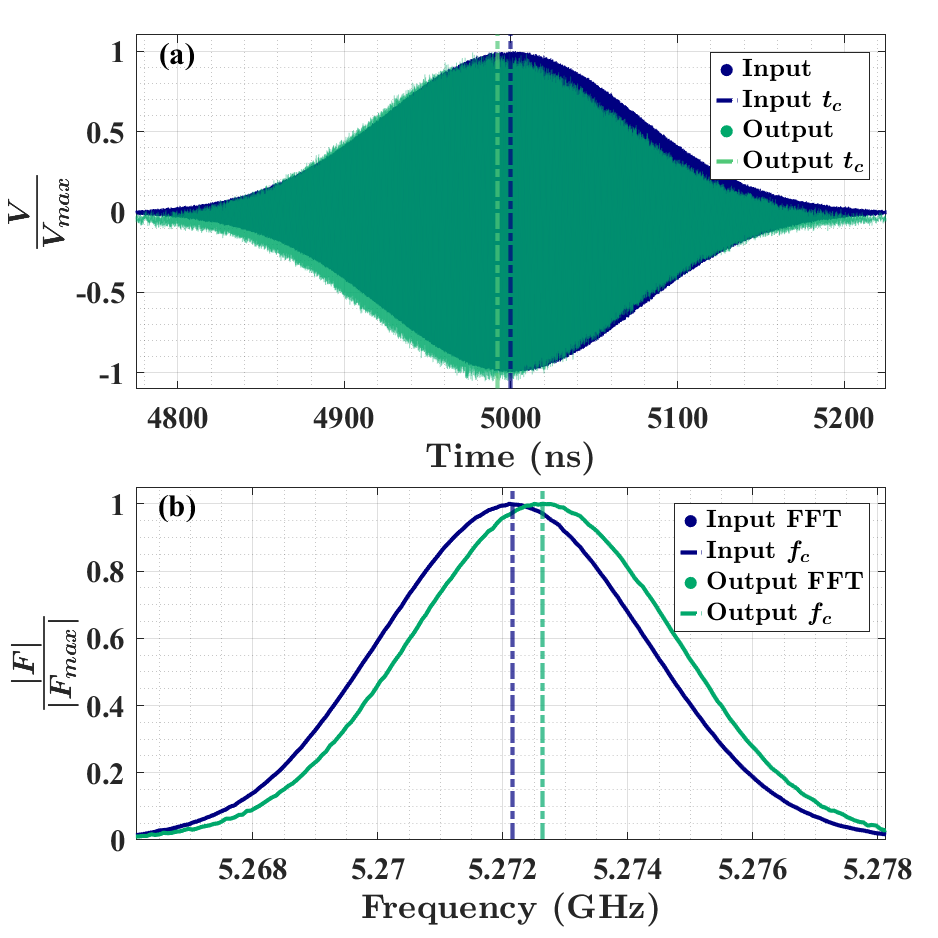}
    \caption{\rev{(a) Example of normalized time domain data for the pulse transmission experiments. The dark blue trace is the pulse that is sent into the ring graph. This pulse has a center frequency of 5.2721 GHz and a frequency bandwidth of 5 MHz. The green trace is the output pulse from the ring graph. Their respective transmission times $t_c$ are plotted as vertical dashed lines. (b) The Fourier transform of the time domain pulse data shown in (a), illustrating the center frequency shift.}} 
    \label{PulseExample}
\end{figure}

In Fig.~\ref{SParam_TransTime}(b) we see that both the real and imaginary parts of the transmission time delay 
evolve through positive and negative values \rev{that can be described in terms of Lorentzian-based functions of frequency \cite{chen2021,chen2022}}. We also see that the transmission time delay extrema coincide with the scattering resonances.

\begin{figure*}[ht!]
    \centering
    \includegraphics[width=\textwidth]{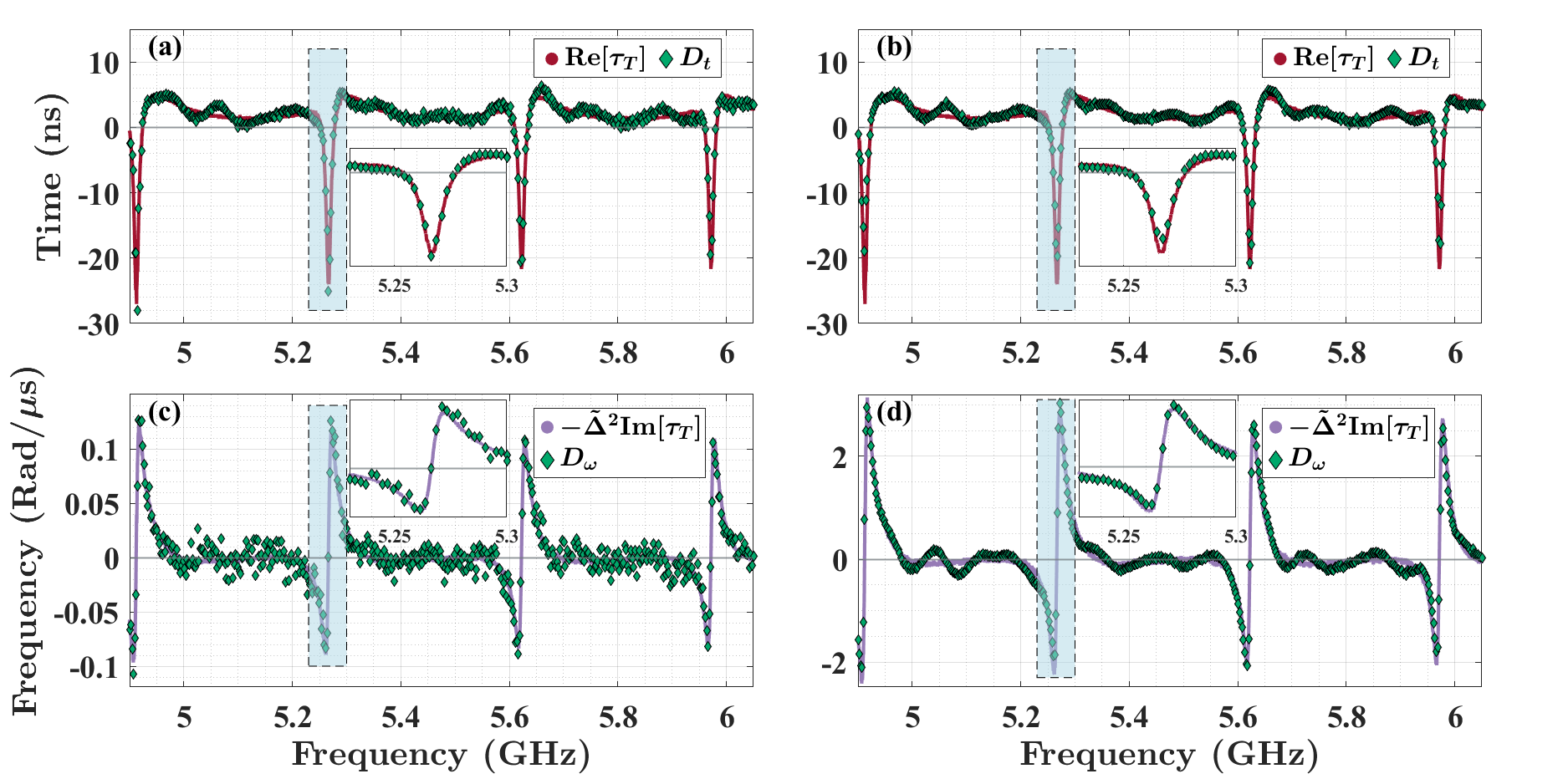}
    \caption{Results for transmission time and center frequency shifts for an input pulse with a frequency bandwidth of 1 MHz ((a) and (c)) and 5 MHz ((b) and (d)). In (a-b) the red curve corresponds to the right side of Eq.~\ref{eq: Dt}. Similarly in (c-d) the purple curve corresponds to the right side Eq.~\ref{eq: Dw}. The green diamonds in plots in the top row (bottom row) are time domain experimental data where $D_t = t_c^{\text{output}}-t_c^{\text{input}}$ ($D_\omega = \omega _c^{\text{output}}-\omega_c^{\text{input}}$) is the difference in the calculated $t_c$ ($\omega_c$) between the input and the output pulses. These correspond to the left hand side of Eqs.~\ref{eq: Dt} and~\ref{eq: Dw}, respectively.}
    \label{mainresults}
\end{figure*}

\textbf{Time Domain Gaussian Pulse Measurements.} The time domain measurements were performed using the setup depicted in Fig.~\ref{exp_schematic}(a). Channel 1 of a 50 GS/s Tektronix model AWG70001B arbitrary waveform generator (AWG) is attached to one end of the ring graph through a coaxial cable. The other end of the ring graph is attached, using another coaxial cable, to channel 1 of a Keysight/Infiniium model UXR0104A 10-GHz bandwidth real-time digital sampling oscilloscope (DSO). The marker channel (M1+) of the AWG is attached to channel 2 of the DSO to trigger the oscilloscope and thus ensure measurements are all taken with the same zero time point. Please see section V in Ref.\cite{SuppMat} for details on how the Gaussian pulses were constructed and how the external delay from the cables was taken into account.

In Fig.~\ref{PulseExample}(a) raw time domain data is shown for both the input and output pulses as well as the measured shifts in time and frequency. 
Note that the oscilloscope measures the detailed carrier-frequency oscillations of the pulse and not just its envelope. The input pulse shown here has a center frequency of \rev{5.2721} GHz which situates it near the center of a resonance of the ring. The frequency bandwidth of the pulse is \rev{5} MHz which is reasonably smaller than the 3-dB bandwidth of this resonance which is about 11.15 MHz. Since we are working in the small bandwidth limit, we calculate the transmission times ($t_c$) and center frequencies ($\omega_c$) using the first temporal moment of the pulse \cite{sebbah1999,Macke03}, defined as,

\noindent
\begin{minipage}{0.45\linewidth}
  \begin{equation}
    t_c = \frac{\int |V(t)|^2t\:dt}{\int |V(t)|^2\:dt}
    \label{t_c}
  \end{equation}
\end{minipage}%
\hspace{0.05\linewidth}
\begin{minipage}{0.5\linewidth}
  \begin{equation}
   \omega_c = \frac{\int |F(\omega)|^2\omega\:d\omega}{\int |F(\omega)|^2\:d\omega}
    \label{w_c}
  \end{equation}
\end{minipage}

\vspace{0.3cm}
\noindent where $V$ is the voltage, $t$ is time, $F$ is the magnitude of the Fourier transform of the time domain signal, and $\omega$ is angular frequency. The deduced transmission times and center frequencies are shown in Fig.~\ref{PulseExample} as vertical lines, demonstrating a negative real time delay of \rev{$D_t = -7.95$ ns and a positive center frequency shift of $D_\omega=3.03$ Rad/$\mu$s or 0.00048 GHz}.


\textbf{Discussion.} A full comparison of the Gaussian pulse measurements in the time domain with the predictions is summarized in Fig.~\ref{mainresults}. The data collected is over 4.9 GHz to 6.05 GHz, including four Feshbach modes, with 480 data points in total taken over the entire frequency range. \rev{(Also see Fig. 5 in the Supp. Mat. \cite{SuppMat} for these results over a broader frequency range (10 MHz to 18 GHz), and Fig. 3 where we explore different pulse frequency bandwidths on a low transmission overlapping mode.)}

We see from Fig.~\ref{mainresults}, that the measured center frequency-shifts $D_\omega$, as well as the measured time shift $D_t$, are in excellent agreement with the predictions of Asano, \textit{et al} (Eqs. \ref{eq: Dt} and \ref{eq: Dw}) \cite{Asano2016}.  These results are also reproduced by simulations of the ring graph (see Section I of \cite{SuppMat}).  Note the difference in scales for the frequency shifts in Figs.~\ref{mainresults} (c) and (d), which shows that the frequency shift of the time-domain pulses increases with the predicted $\tilde{\Delta}^2$ scaling.  \rev{Also note that Figs.~\ref{mainresults} (a) and (b) are nearly identical (i.e. independent of pulse bandwidth), as predicted}. In all cases there are systematic deviations between the time-domain results and the predicted values from frequency-domain complex time delay in the range between the Feshbach modes. These deviations are attributed to standing waves on the input and output coaxial cables used in the time-domain measurements (see Ref.~\cite{SuppMat} sections I, II, and VI for more details). 

One interesting observation is that the imaginary part of complex time delay produces changes in the carrier frequency so as to decrease the amount of absorption of the transmitted pulse \cite{garrett1970,Macke03}. Related to this, Ref. \cite{Talukder01} show a clear deviation from exponential decrease of laser intensity with propagation distance in a dispersive absorbing medium, showing that the light is less attenuated at greater distances than one would expect.  In the case of a scattering system with gain, it has also been noted that the center frequency shift will be towards (rather than away from) the gain mode \cite{garrett1970,Macke03}. \rev{It is also worth pointing out that there is a clear correspondence between where negative time delay occurs and where the pulse center-frequency shifts are present. The physical mechanism behind these frequency shifts is the same as that giving rise to negative time delays, where they are a result of nonuniform distortion of the Fourier components of the pulse as it travels through a dispersive medium \cite{boyd2002, garrett1970,gehring2006}.} 

Asano, \textit{et al} also make predictions for the maximum time and frequency shifts that can be created by a given scattering system \rev{in the critical coupling limit}. These \rev{upper bounds are} analogous to those for expectation values in quantum weak measurements,\cite{Asano2016} and superoscillatory functions \cite{Berry06,Berry09}.  The bounds on time and frequency shifts are given by, $D_{\text{t,max}} = \pm \frac{1}{\sqrt{2}\tilde{\Delta}}$ and $D_{\omega,max} = \pm \frac{\tilde{\Delta}}{\sqrt{2}}$, respectively. In our case, for the pulses with a 1 MHz bandwidth, this would result in  $D_{\text{t,max}} \approx 265$ ns and $D_{\omega,max} \approx 12$ Rad/$\mu$s, while for the 5 MHz bandwidth pulse case one has  $D_{\text{t,max}} \approx 53$ ns and $D_{\omega,max} \approx 59$ Rad/$\mu$s. Our data for both of these cases, presented in Fig. \ref{mainresults} (as well as Fig. 5 in Section IV of Ref.~\cite{SuppMat}), are clearly well within these bounds, \rev{which is expected because the graph measurement is in the strong-coupling limit}.

Our work generalizes that of Asano, \textit{et al}. \cite{Asano2016} in the sense that our results for $D_t$ and $D_{\omega}$ are not tied to any particular model of transmission near a resonant mode.  We have shown \rev{instead} that complex time delay derived from frequency-domain data provides model-free predictions for the pulse modifications due to scattering. 
\rev{We have shown that} this includes frequencies that are far from resonant modes, where the analytical approximations are no longer valid.

\textbf{Conclusions.} In this paper we experimentally demonstrate the connection between complex transmission time delay and Gaussian pulse properties; verifying the predictions first laid out in Ref. \cite{Asano2016}. The most novel contribution is the direct connection between the imaginary component of the transmission time delay and the center frequency shift of the scattered Gaussian pulse. This helps bring physical meaning to an abstract but practically useful quantity that makes up the complex time delay.

\rev{This paper establishes the detailed equivalence of complex scattering information derived from frequency-domain and time-domain approaches, providing insights that inform and simplify measurements over the entire electromagnetic spectrum.}


\rev{In terms of future work, it would be interesting to generalize these predictions to arbitrary pulse shapes. It would also be interesting to see how this relation would hold for more complex scattering systems with overlapping modes, as well as for gain modes, or systems with strong nonlinearities.} Additionally, we can now make predictions for reflection time delays, along with reflection time-delay differences \cite{fyodorov2019,osman2020}, as well as transmission time-delay differences in non-reciprocal scattering systems \cite{Nadav24}. The connection of this work to extreme time delays associated with scattering singularities \cite{Nadav24,Erb2025} is also of interest.

\begin{acknowledgments}
 We thank Nadav Shaibe for helpful discussions.  This work was supported by NSF/RINGS under grant No. ECCS-2148318, ONR under grant N000142312507, and DARPA/WARDEN under grant HR00112120021.
\end{acknowledgments}

\newpage
\bibliography{Main}
\end{document}


\preprint{APS/123-QED}

\title{A Physical Interpretation of Imaginary Time Delay\\
Supplementary Material}

\author{Isabella L. Giovannelli}
 \email{igiovann@umd.edu}
\affiliation{%
Maryland Quantum Materials Center, Department of Physics\\
 University of Maryland, College Park, Maryland 20742, USA
}%

\author{Steven M. Anlage}
\affiliation{%
Maryland Quantum Materials Center, Department of Physics\\
 University of Maryland, College Park, Maryland 20742, USA
}%

\date{\today}

\maketitle


Here we discuss further details about numerical simulations of the ring graph in both the frequency-domain and time-domain (Section \ref{Sims}), the renormalization of complex time delay for low-transmission (Section \ref{LowT}), details of the transmitted pulse analytical calculation (Section \ref{TPulseCalc}), plots of the data over the full measurement frequency range (Section \ref{DataPlots}), detailed discussion of the time-domain measurements (Section \ref{TDMeas}), a discussion of both systematic and random errors in the experiment (Section \ref{Errors}), and further background information about complex time delay (Section \ref{BkdCTD}).

\section{Simulation} \label{Sims}
\begin{figure}[ht!]
    \centering
    \includegraphics[width=0.25\textwidth]{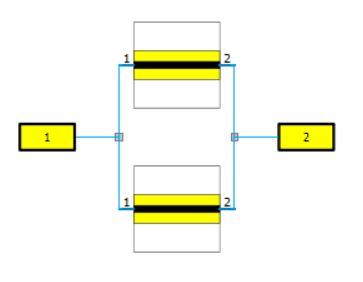}
    \caption{CST model of microwave ring graph used for simulations consisting of two ports, two T-junctions (small gray squares), and two coaxial cables (large gray squares). The coaxial cables are identical except for their lengths which differ by 1 inch. The blue lines are zero-length electrical connections.}
    \label{CST_model_1}
\end{figure}

The frequency-domain and time-domain simulations of the ring graph were performed using CST Studio Suite 2021. Fig.~\ref{CST_model_1} shows the model used for the CST simulations.

This model consists of two ports, two T-junctions, and two coaxial cables. The diameters of the inner and outer conductors for the coaxial cables are 0.091 cm and 0.298 cm respectively. The other parameters set for the coaxial cables include the relative permittivity ($\epsilon_r$) which is 2.01047, the dielectric loss tangent which is $\tan \delta = 0.00028$, and the metal resistivity normalized to gold resistivity which is 1.8. The upper coaxial cable in Fig.~\ref{CST_model_1} has a length of 11 in (27.9 cm) and the lower coaxial cable has a length of 12 in (30.5 cm).

The ring graph is interesting because it supports two types of modes \cite{Waltner13,Waltner14}.  One set have standing wave maxima near the leads, establishing strong coupling to the leads, and creating low-Q resonances known as shape resonances.  The other modes are standing wave patterns rotated by 90-degrees in phase compared to the shape modes, leading to smaller coupling to the leads and high-Q resonances, and are known as Feshbach resonances.  The features arising from each of these two classes of resonances are evident in the $S_{21}(f)$ data shown in Figs.~2(a) and 4(a) below.

\begin{figure}[ht!]
    \centering
    \includegraphics[width=0.4\textwidth]{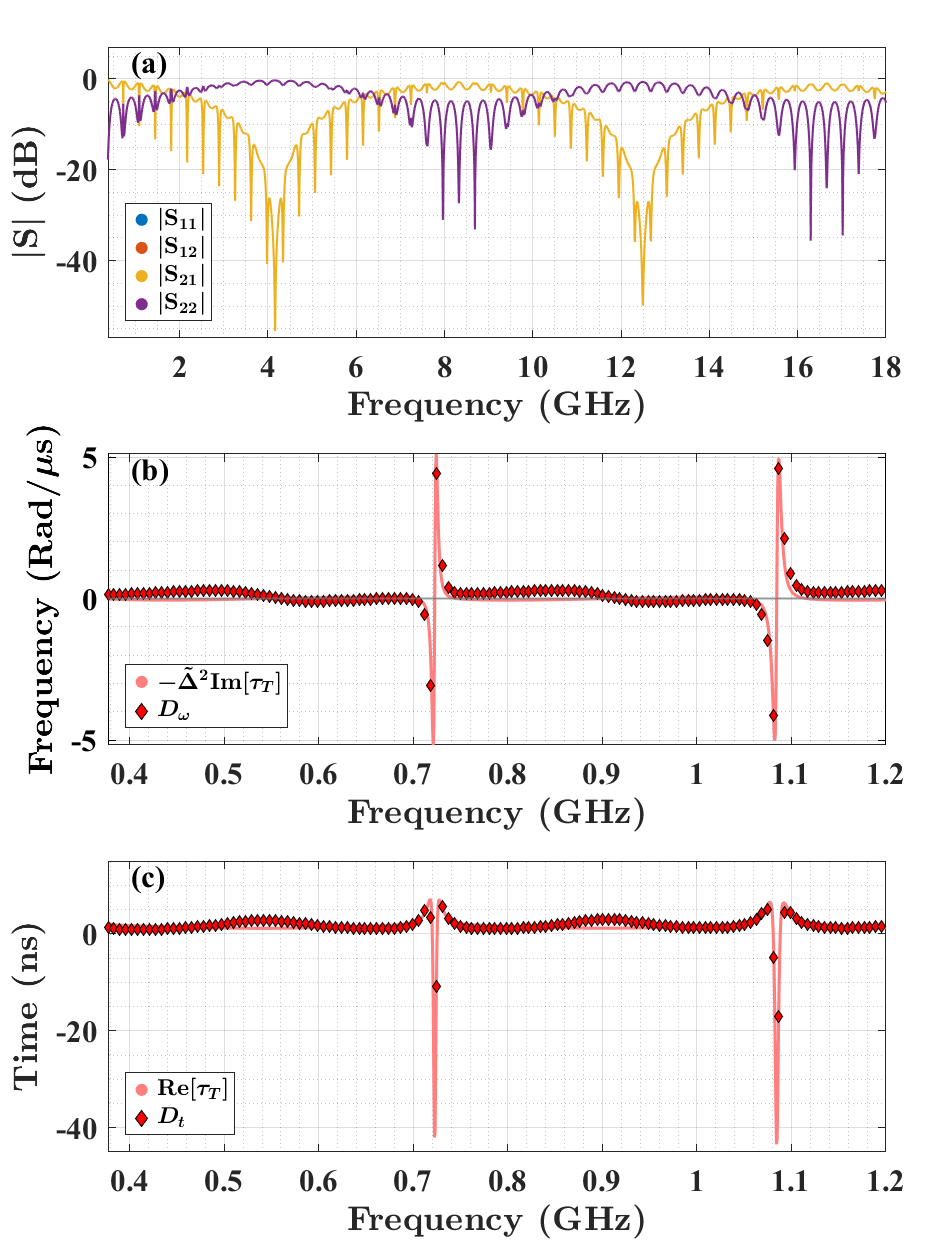}
    \caption{(a) Simulated scattering parameters for the ring graph model depicted in Fig.~\ref{CST_model_1}. (b) Simulation results for the frequency shift of the Gaussian pulse with a bandwidth of 5 MHz. (c) Corresponding simulation results for the time shift of a Gaussian pulse with a bandwidth 5 MHz.}
    \label{simulation_summary}
\end{figure}

Figure \ref{simulation_summary} contains a summary of our main simulated results. Fig.~\ref{simulation_summary}(a) are the four scattering parameters of the 2-port device as a function of frequency, which we note are very similar in character to those for the measured ring graph shown in Fig.~\ref{S_param_full}(a) below. Note that these results are obtained using the frequency domain ``S-parameters" task in CST. The complex transmission time delay is calculated from $S_{21}(f)$.  In Fig.~\ref{simulation_summary}(b),  Eq. 3 from the main text is plotted for the case of a 5 MHz bandwidth pulse. The shift in center frequency of the simulated Gaussian time-domain pulse ($D_\omega$) is plotted on top as red diamonds. This is done over the center-frequency range of 0.377 GHz to 1.2 GHz for a 5 MHz bandwidth Gaussian pulse. Note that the Gaussian pulse simulations are done using the time domain ``transient" task in CST where the input excitations (i.e. Gaussian pulses) are generated in Matlab using the same equation used in the experiment (see Eq.\ref{TD_E_eq}).

\begin{figure*}[t!]
    \centering
    \includegraphics[width=\textwidth]{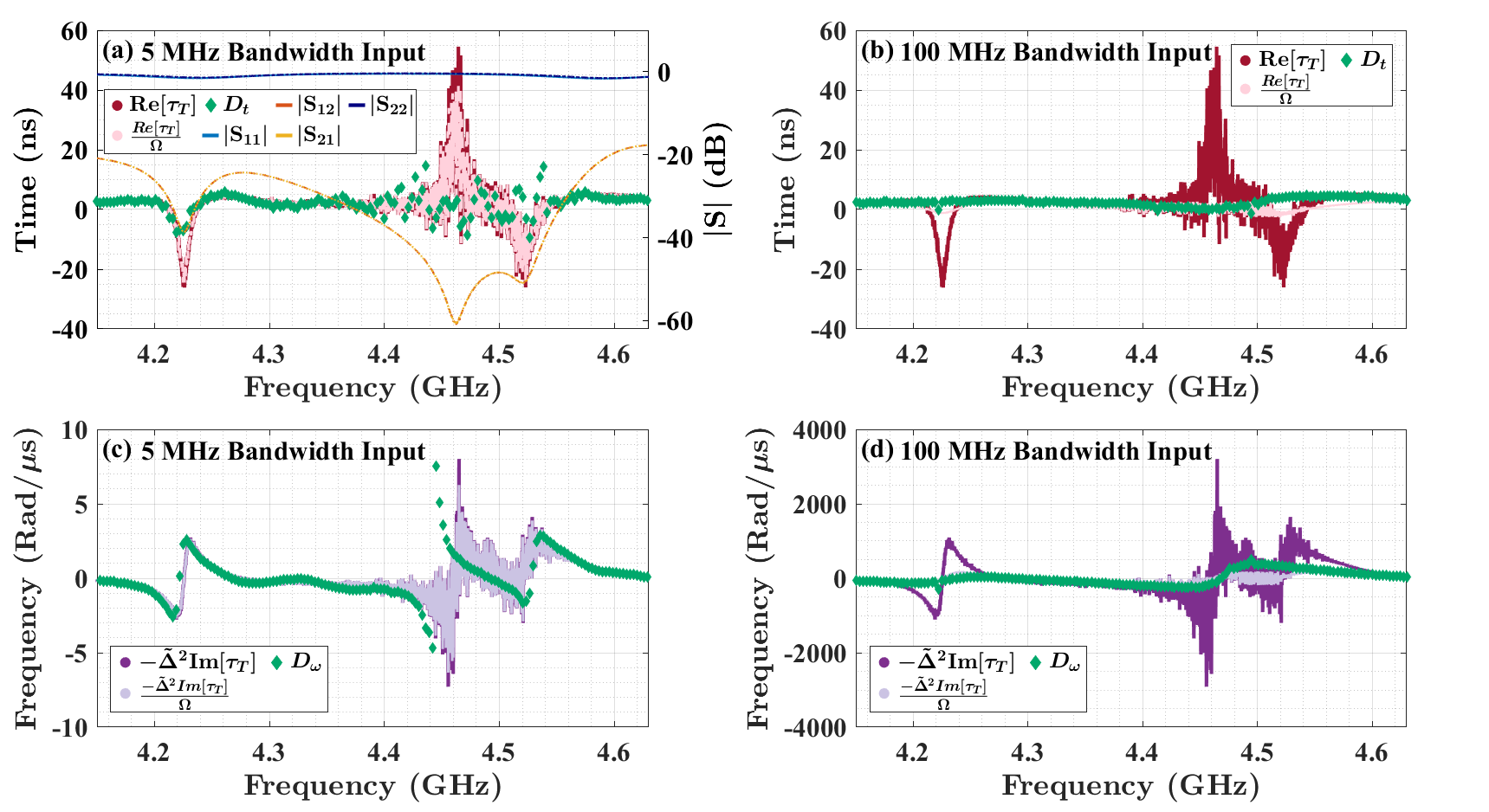}
    \caption{Summary of low transmission and large pulse bandwidth results. (a) This plot is for the case where the frequency bandwidth of the input pulse is 5 MHz. The left axis is plotting the shift in time. The real part of the transmission time delay is plotted in dark red (Re[$\tau_T$]) and the adjusted form of Re[$\tau_T$] for extreme low transmission is plotted in pink where $\Omega = 1+\frac{1}{2}\tilde{\Delta}^2|\tau_T|^2$ is the correction factor. The shift in transmission time of the Gaussian pulses is shown as green diamonds ($D_t$). The right axis is in dB and corresponds to the scattering parameters listed in the legend. Note that these scattering parameters are the same for every plot in this figure. (b) Same as (a) except this plot is for the case where the frequency bandwidth of the input pulse is 100 MHz. (c-d) These are the corresponding plots for the shift in frequency case. In dark purple is the imaginary component of transmission time delay scaled by the bandwidth factor (-$\tilde{\Delta}^2$Im[$\tau_T]$). On top in light purple is -$\tilde{\Delta}^2$Im[$\tau_T]$ corrected for the case of extreme low transmission. The green diamonds are the measured shift in frequency of the Gaussian pulses sent through our experiment ($D_\omega$).}
    \label{low_transmission_example}
\end{figure*}

Fig.~\ref{simulation_summary}(c) shows the time shift predictions, analogous to Fig.~\ref{simulation_summary}(b), as a function of pulse center-frequency. The real part of the transmission time delay is obtained directly from the frequency domain simulation of $S_{21}(f)$. The shift in transmission time of the simulated time-domain Gaussian pulse ($D_t$) is plotted on top as red diamonds. This is done over the same center-frequency range of 0.377 GHz to 1.2 GHz for a 5 MHz bandwidth Gaussian pulse.  The agreement between the two methods of calculating center frequency shift and time delay agree to the same extent as in the experiment.

\section{Extreme Low Transmission and Large Pulse Bandwidth Limits}  \label{LowT}
It was noted in Eq. 11 of Ref.~\cite{Asano2016} that in the extreme low transmission limit one must adjust the predictions presented in Eqns. 2-3 in the main text. For convenience, these predictions are reproduced below in the notation used in this paper,
\begin{align}
    D_{t} = \frac{\text{Re}[\tau_T]}{\Omega} \qquad D_{\omega} = \frac{-\tilde{\Delta}^2\text{Im}[\tau_T]}{\Omega}
    \label{RenormEq}
\end{align}
where again $\tau_T$ is the transmission time delay defined in equation 1 of the main text, $\tilde{\Delta}=\frac{\tilde{\Delta}_\omega}{2\sqrt{2\text{ln}2}}$ where $\tilde{\Delta}_\omega$ is the angular frequency bandwidth of the pulse. The renormalization term $\Omega$ is defined as $\Omega=1+\frac{1}{2}\tilde{\Delta}^2|\tau_T|^2$. The addition of this renormalization term prevents divergence of the predicted time delay and frequency shift.

In Figure \ref{low_transmission_example} we show a summary of results in the limits of low transmission and large pulse bandwidths. These measurements were performed over the mode with the lowest level of transmission in our system, which is in the frequency range 4.15 GHz to 4.63 GHz. To see how the transmission at this mode compares with others in the system, please see Fig.~\ref{S_param_full}(a) for a plot of the scattering parameters over the frequency range 0.377 GHz to 18 GHz. 

For the small pulse bandwidth case of 5 MHz (Fig.~\ref{low_transmission_example}(a) and (c)) we see that low-transmission predictions (Eqn. \ref{RenormEq}) are both smaller in amplitude than the original predictions (Eqns. 2-3 of the main text). Here we see that the time domain experiment results seem to better follow the corrected low-transmission predictions in that the peaks are less extreme. Note that in this region the transmission is quite low ($< -40\ dB$) and the modes in the 4.4 GHz - 4.55 GHz region are non-trivial and overlapping (as seen in the associated scattering parameters plotted in Fig.~\ref{low_transmission_example}(a)). This made this case particularly difficult to examine experimentally and is why there is a substantial amount of noise present. 

We do not see much difference between the low transmission and original predictions until a Gaussian pulse with a large bandwidth is considered. These results are shown in Fig.~\ref{low_transmission_example}(b) and (d) for a pulse with a bandwidth of 100 MHz. Here we see an extreme decrease in the amplitude of the predictions (Eqn. \ref{RenormEq}) relative to Eqns. 2-3 of the main text. The time domain experimental data (in green) clearly follows the corrected low-transmission predictions more closely than the original predictions.

Overall, we see that the renormalized predictions for frequency and time shifts in low-transmission regions are generally valid. Particularly for the case where the pulse bandwidth is large in comparison to the 3 dB bandwidth of the chosen resonance. For the small bandwidth case, the experimental results are still consistent with the predictions, but the difference that the correction makes is negligible. 

\section{Transmitted Pulse Calculation Details}  \label{TPulseCalc}
In this section we reproduce the derivation linking transmission time delay to the characteristics of a Gaussian pulse. This is done by combining methods used in Asano \textit{et al}. \cite{Asano2016} and Cao \textit{et al}. \cite{cao2003}.  The assumptions underlying this calculation include the following : 1) The frequency bandwidth of the pulse $\tilde{\Delta}_\omega$ is much smaller than the  3-dB linewidth of the resonant mode being studied $\gamma_{3-\text{dB}}$, and 2) \rev{the system is linear and dispersive.}

The frequency domain equation for the input Gaussian pulse is given by:
\begin{align}
    E_i(\omega)=\frac{B}{\tilde{\Delta}}\text{exp}\left[-\frac{(\omega-\omega_c)^2}{2\tilde{\Delta}^2}\right]
    \label{eq: E_in}
\end{align}
where $\omega_c$ is the carrier angular frequency, $B$ is the initial amplitude, and 
\begin{align}
    \rev{\tilde{\Delta}=\frac{\tilde{\Delta}_\omega}{2\sqrt{2\text{ln}2}}}
    \label{eq: bandwidth}
\end{align}
where $\tilde{\Delta}_\omega$ is the desired angular frequency bandwidth of the pulse. \rev{Note that $\tilde{\Delta}_\omega$ is equal to the full width at half maximum (FWHM) of the Gaussian curve in the frequency domain space. See section V for more details.}

The frequency domain equation for the output Gaussian pulse is given by:
\begin{align}
    E_o(\omega;\alpha) = T(\omega + i\alpha)E_i(\omega)
    \label{eq: E_out}
\end{align}
where $T(\omega+ i \alpha)$ is the transmission coefficient of the lossy scattering system, and $\alpha$ quantifies the uniform loss. Following the transcription in \cite{cao2003}, we define 
\begin{align}
   A(\omega;\alpha)=\frac{1}{|T(\omega+i\alpha)|}
   \label{eq: A}
\end{align}
which quantifies both absorption (when $|T(\omega+i\alpha)|<1$) and gain (when $|T(\omega+i\alpha)|>1$). We then define the real ``transfer coefficient" to be $\kappa(\omega;\alpha) = \text{ln}(A(\omega;\alpha))$. Putting this together we get,
\begin{align}
    E_o(\omega;\alpha) = \text{exp}[-\kappa (\omega;\alpha) + i \phi(\omega;\alpha)]\: E_i(\omega)
    \label{eq: E_out_2}
\end{align}
where $\phi$ is the phase of the transmission coefficient.

We now require that the bandwidth of this Gaussian pulse to be narrow such that $|T(\omega;\alpha)|$ and $\phi(\omega;\alpha)$ only experience slight variation over the bandwidth of the pulse. This then allows us to expand $\kappa(\omega;\alpha)$ and $\phi(\omega;\alpha)$ into a Taylor series, centered around the carrier frequency $\omega_c$
\begin{align}
    \kappa(\omega;\alpha) = \kappa(\omega_c;\alpha) + \left.\frac{d\kappa(\omega;\alpha)}
    {d\omega}\right|_{\omega_c}(\omega-\omega_c)+\cdots  \label{eq: kappa}\\
    \phi(\omega;\alpha) = \phi(\omega_c;\alpha) + \left.\frac{d\phi(\omega;\alpha)}{d\omega}\right|_{\omega_c}(\omega-\omega_c)+\cdots \label{eq: phi}
\end{align}
in this case we assume the bandwidth is sufficiently narrow such that higher order terms can be neglected. These expansions can then be related to the real and imaginary parts of the transmission time delay. Calculating the transmission time delay associated with the generalized transmission coefficient $T(\omega+i\alpha)=\text{exp}[-\kappa (\omega;\alpha) + i \phi(\omega;\alpha)]$ yields, 
\begin{align}
    \tau_T(\omega;\alpha)&=-i\frac{\partial}{\partial\omega}\text{ln}\left[\text{exp}[-\kappa (\omega;\alpha) + i \phi(\omega;\alpha)]\right] \label{eq: derivative_T_1}\\
    &= \frac{\partial\phi(\omega;\alpha)}{\partial\omega}+i\frac{\partial \kappa(\omega;\alpha)}{\partial \omega} \label{eq: derivative_T_2}\\
    &= Re[\tau_T]+iIm[\tau_T] \label{eq: derivative_T_3}
\end{align}
Combining equations \ref{eq: kappa} and \ref{eq: phi} with \ref{eq: derivative_T_2} and \ref{eq: derivative_T_3} results in the following expressions for $\kappa$ and $\phi$,
\begin{align}
    \kappa(\omega;\alpha) \approx \kappa(\omega_c;\alpha) + \text{Im}[\tau_T(\omega_c;\alpha)](\omega-\omega_c) \label{eq: kappa2}\\
    \phi(\omega;\alpha) \approx \phi(\omega_c;\alpha) + \text{Re}[\tau_T(\omega_c;\alpha)](\omega-\omega_c) \label{eq: phi2}
\end{align}
Putting these results (lines \ref{eq: kappa2} and \ref{eq: phi2}) into equation \ref{eq: E_out_2} results in the following,
\begin{widetext}
    {\small
    \begin{align*}
     E_o(\omega;\alpha) = &\: E_i(\omega)\times\text{exp}[-(\kappa + \text{Im}[\tau_T](\omega-\omega_c))]\times\text{exp}[i (\phi + \text{Re}[\tau_T](\omega-\omega_c))] \\
     = &\: E_i(\omega)T(\omega_c+i\alpha)\times\text{exp}[(-\text{Im}[\tau_T]+i\text{Re}[\tau_T])(\omega-\omega_c)] \\
     &\text{Now we add in the expression for the Gaussian pulse for the input wave (Eqn. \ref{eq: E_in}).}\notag \\
     = &\: \frac{B}{\tilde{\Delta}}\:T(\omega_c+i\alpha)\times\text{exp}\left[-\frac{(\omega-\omega_c)^2}{2\tilde{\Delta}^2}-\text{Im}[\tau_t](\omega-\omega_c)\right]\times\text{exp}\left[i\text{Re}[\tau_T](\omega-\omega_c)\right] \\
      &\text{Next we complete the square in the argument of the exponential.}\notag \\
     = &\: \frac{B}{\tilde{\Delta}}\:T(\omega_c+i\alpha)\times\text{exp}\left[\frac{-\left((\omega-\omega_c)+\tilde{\Delta}^2\text{Im}[\tau_T]\right)^2}{2\tilde{\Delta}^2}+\frac{1}{2}\tilde{\Delta}^2\text{Im}[\tau_t]^2\right]\times\text{exp}[i\text{Re}[\tau_T](\omega-\omega_c)] \\
     = &\: \frac{B}{\tilde{\Delta}}\:T(\omega_c+i\alpha)C_o(\omega_c;\alpha)\times\text{exp}\left[\frac{-\left((\omega-\omega_c)-D_\omega\right)^2}{2\tilde{\Delta}^2}\right]\times\text{exp}[iD_t(\omega-\omega_c)] \\
     = &\: T(\omega_c+i\alpha)C_o(\omega_c;\alpha)\times E_i(\omega-D_\omega)\times\text{exp}[iD_t(\omega-\omega_c)] 
     \end{align*}
    }
\end{widetext}
where $C_o(\omega_c;\alpha) = \text{exp}\left[\frac{1}{2}\tilde{\Delta}^2\text{Im}[\tau_T(\omega_c;\alpha)]^2\right]$ is constant at a specific center frequency $\omega_c$ and loss parameter $\alpha$. From this final expression, we see that the output pulse is shifted in the frequency space by $D_\omega=-\tilde{\Delta}^2\text{Im}[\tau_T(\omega;\alpha)]$, relating the shift to the imaginary part of the transmission time delay. In the time domain, one can show in an analogous fashion that the temporal shift of the pulse is $D_t=\text{Re}[\tau_T(\omega;\alpha)]$, which relates the time-shift of the pulse to the real part of the transmission time delay.

\section{Plots of data over the full frequency range}  \label{DataPlots}
\begin{figure}[ht!]
    \centering
    \includegraphics[width=0.48\textwidth]{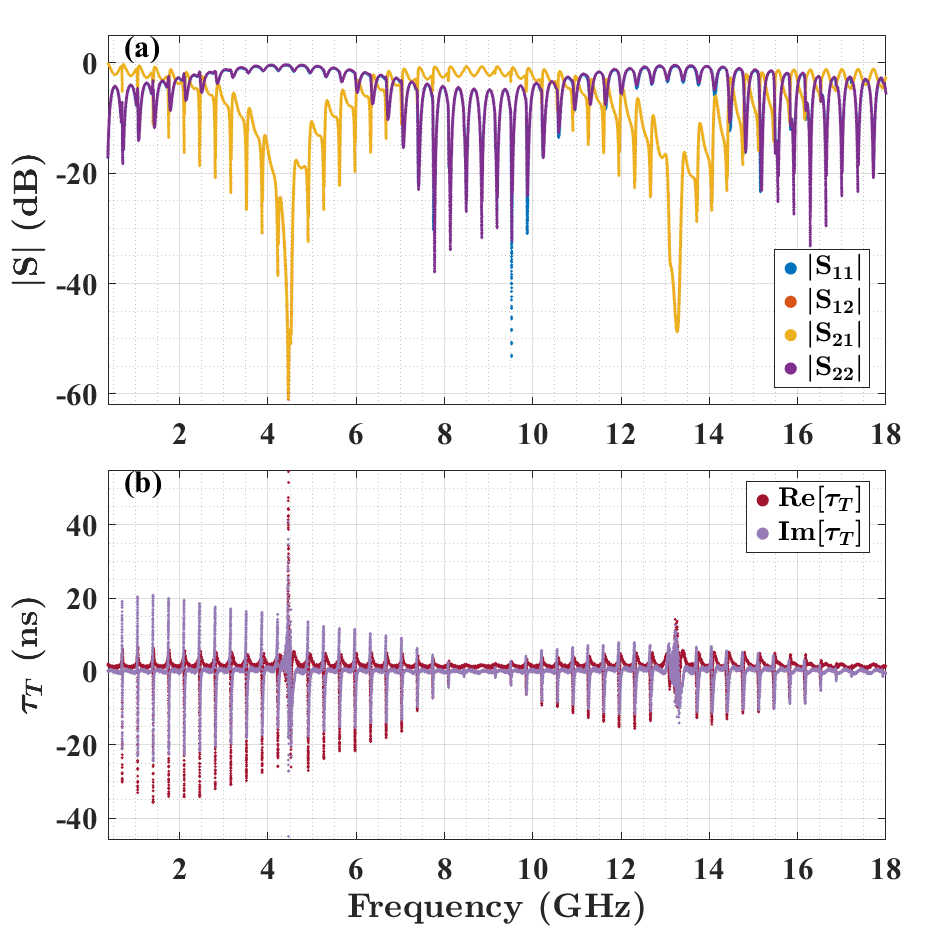}
    \caption{(a) Full S-parameter data for the microwave ring graph collected over the frequency range 0.377 GHz to 18 GHz. (b) The corresponding calculated transmission time delay $\tau_T$ (in ns) with both its real and imaginary parts plotted.}
    \label{S_param_full}
\end{figure}

In this section we show the measured S-parameters and complex time delay of the ring graph over the entire measured frequency range (0.377 - 18 GHz). Figure~\ref{S_param_full}(a) shows the scattering parameters for the ring graph shown in Fig.~1 of the main text. These are measured using a calibrated network analyzer. See the main text `Experiment' section for more details.   Figure \ref{S_param_full}(b) shows the derived complex transmission time delay (in ns) as a function of frequency for the ring graph.

\begin{figure*}[ht!]
    \centering
    \includegraphics[width=\textwidth]{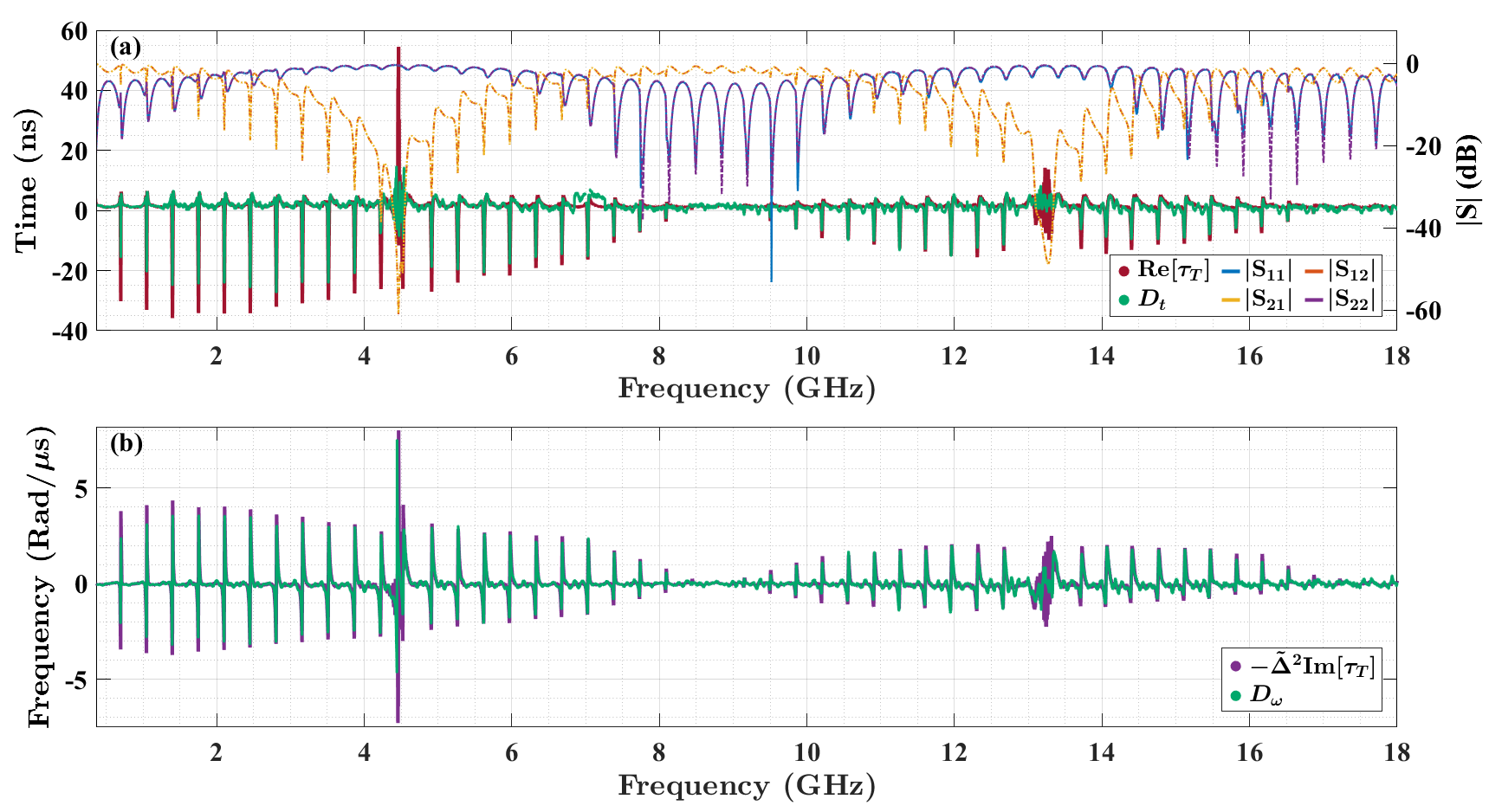}
    \caption{Summary of the main results for the case where the input Gaussian pulse has a 5 MHz bandwidth. The data was collected over the frequency range 0.377 GHz to 18 GHz. (a) shows the time delays Re$[\tau_T]$ and $D_t$ as a function of frequency, and (b) shows the frequency shifts $-\tilde{\Delta}^2$Im[$\tau_T$] and $D_\omega$ as a function of frequency. In (a) the full set of scattering parameters are plotted using the right axis.}
    \label{full_data}
\end{figure*}

In this section we also have Gaussian pulse time and frequency shift results over the full frequency range from 0.377 GHz to 18 GHz. These results are shown in Fig.~\ref{full_data} where Fig.~\ref{full_data}(a) illustrates the time shift results and Fig.~\ref{full_data}(b) shows the analogous frequency shift results. The scattering parameters are reproduced on each plot to show the correspondence between the resonant modes and the experienced time/frequency shifts. In both cases we see that in general we have excellent agreement between $D_t$ and Re$[\tau_T]$, and $D_\omega$ and $-\tilde{\Delta}^2$Im[$\tau_T$]. In other words, this data indicates that there is a strong connection between complex time delay and the center time and frequency shifts experienced by a Gaussian pulse as it travels through a ring-graph resonator. Specifically, this supports the conclusion that the real part of transmission time delay corresponds to a time shift in the Gaussian pulse and the imaginary part of transmission time delay is related to a center frequency shift in the Gaussian pulse.

In both Fig.~\ref{full_data}(a) and Fig.~\ref{full_data}(b) we see that as frequency increases, there is an increase in oscillatory behavior in the green time domain data ($D_t$ and $D_\omega$) in the regions in between the Feshbach modes. For example, in the region 0.377 GHz to 0.6 GHz, we see that the green time domain data is mostly flat and in agreement with the predictions (Eqns. 2-3 in the main text). However, in the region around 12 GHz to 12.25 GHz, we see that the green time domain data has picked up some oscillatory behavior that is not present in the time delay curves (in red and purple). We believe that this is a result of a discrepancy between the system used for time domain measurements and the frequency domain setup. Please see section VI of the Supp. Mat. for more details.

\begin{figure*}[ht!]
    \centering
    \includegraphics[width=\textwidth]{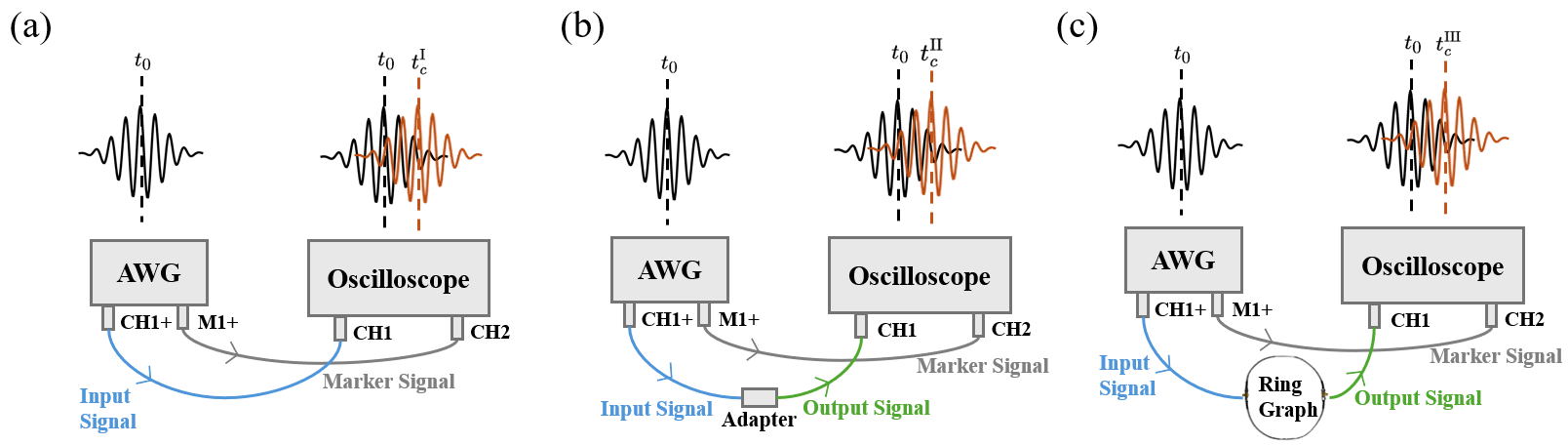}
    \caption{Summary of the different time domain setups used to measure and remove the time delay associated with the input and output cables, and to determine the delay associated with just the ring graph. (a) Schematic of the time delay measurement to determine delay associated with the input cable (blue). (b) Schematic of the time delay measurement to determine delay associated with both the input cable and output cable (green) along with a Female-Female adapter. (c) Schematic of the time delay measurement to determine delay associated with the input cable, ring graph, and output cable.}
    \label{time_delay_subtraction}
\end{figure*}

\section{Time domain pulse measurements}  \label{TDMeas}
In this section we go into more detail about how the Gaussian pulses are created in the experiment. We also discuss how time delay and center-frequency shift are measured in the time domain setup. Specifically here we discuss how the time delay arising from the input and output cables (shown in Fig.~6(c)) is removed from the measurement to isolate the time delay just due to the graph.

\subsection{Defining Gaussian pulse and $\tilde{\Delta}$}
The Gaussian pulses used in our experiment are created on the AWG using the equation,
\begin{align}
    \rev{E_i(t) = \text{cos}(\omega_c (t-t_0)) \text{exp}\left[-\frac{(t-t_0)^2\tilde{\Delta}^2}{2}\right]}
    \label{TD_E_eq}
\end{align}
where $\omega_c = 2\pi f_c$ and $f_c$ is the selected center frequency of the Gaussian pulse. The variable $t_0$ is where the pulse is centered in time in the AWG memory, and this is chosen to be 3000 ns for the 1 MHz bandwidth case and 5000 ns for the 5 MHz bandwidth case.  These values are chosen so that the pulse is fully generated by the AWG. Note that this choice is mostly arbitrary and the exact number itself does not hold any physical significance. 

\rev{Lastly, $\tilde{\Delta}$ (defined in Eqn. \ref{eq: bandwidth} above) was originally defined as the frequency bandwidth of the Gaussian pulse in reference \cite{Asano2016}. We however found this definition to be inadequate and had to redefine $\tilde{\Delta}$ to be in terms of the full width at half maximum (FWHM) of the Gaussian distribution of the pulse in frequency space. First we assume that $\tilde{E}(\omega)$ takes the form of Eqn. 2 in \cite{Asano2016}, reproduced below for convenience,
\begin{align}
    \tilde{E}(\omega)\propto\text{exp}\left[-\frac{(\omega-\omega_c)^2}{2\tilde{\Delta}^2}\right]
    \label{eq: DeltaDeriv1}
\end{align}
To find the FWHM of this distribution we perform the following calculation,
\begin{align}
    \frac{1}{2}\tilde{E}_{\text{max}}&=\tilde{E}_{\text{max}}\text{exp}\left[-\frac{(\omega_\text{HM}-\omega_c)^2}{2\tilde{\Delta}^2}\right] \\
    \omega_\text{HM} &=\pm\tilde{\Delta}\sqrt{2\text{ln}2}+\omega_c
    \label{eq: DeltaDeriv1}
\end{align}
where $\omega_\text{HM}$ is the frequency at half maximum. The FWHM is then,
\begin{align}
    \Delta_\omega=\Delta\omega_\text{HM} &=2\tilde{\Delta}\sqrt{2\text{ln}2}
    \label{eq: DeltaDeriv2}
\end{align}
Thus, we now have $\tilde{\Delta}$ in terms of the FWHM ($\Delta_\omega$) of the Gaussian distribution of the pulse in frequency space.}

\subsection{Removing extraneous time delays}
In all of our measurements we are comparing the properties of a Gaussian pulse that has traveled through the ring graph (called the output pulse) to those of the original input Gaussian pulse. In both of these cases there is additional delay due to the cables in our time-domain system (external to the ring graph) that has to be removed. See Fig.~\ref{time_delay_subtraction} for a visualization of these cases.  Note that these additional cables do not play a role in the frequency domain measurements because they are calibrated out of the S-parameter measurement.

For the input pulse case, first the Gaussian pulse is created using the AWG and centered at a particular time $t_0$ in the device memory. The pulse is output from the AWG and sent through the same cable that would normally be connected to the ring graph, except in this case the cable to connected directly to the oscilloscope where the input pulse (that will delivered to the ring-graph) can then be measured, see Fig.~\ref{time_delay_subtraction}(a). The transmission time $t_c^{\text{I}}$ is calculated from the measured pulse in this case using the same formula defined in the main text in Eq.~4. We know that if there were no delay in the cable the measured transmission time of the pulse should be $t_0$ in the oscilloscope memory, so we subtract out any additional delay added by the input cable ensuring the measured input pulse is centered at $t_0$. Thus, $\Delta t_{delay}^{in}=t_c^{\text{I}} - t_0$ where we now define $t_c^{input} = t_c^{\text{I}} - \Delta t_{delay}^{in}=t_0$.

A similar process is performed for the output pulse case. For this case, the pulse first travels through the input cable, through the ring graph, and then through an output cable. See Fig.~\ref{time_delay_subtraction}(c). To remove the time delay in the cables we perform a measurement where we remove the ring graph and replace it with an electrically short Female-to-Female (f2f) coaxial cable adapter, see Fig.~\ref{time_delay_subtraction}(b). From here we repeat the same process as for the input pulse case. We subtract any delay ensuring that the output pulse is centered at $t_0$ in the oscilloscope memory. The additional delay from the f2f adapter is taken into account by measuring its electrical length using a microwave vector network analyzer and then estimating its delay. The estimated delay for this f2f adapter is approximately $t_{delay}^{adapter}=0.014$ ns. The total delay for this case is calculated as $\Delta t_{delay}^{out}=t_c^{\text{II}} - t_0 - t_{delay}^{adapter}$ where $t_c^{\text{II}}$ is the time depicted in Fig.~\ref{time_delay_subtraction}(b) 
 and is calculated using Eq.~4 in the main text. Finally we calculate $t_{c}^{output} = t_c^{\text{III}} - \Delta t_{delay}^{out}$ for the final transmission time value with the delay arising from the cables removed. The time $t_c^{\text{III}}$ is the transmission time measured in the setup of Fig.~\ref{time_delay_subtraction}(c). Note that the $D_t$ values shown in the main text are then calculated as $D_t = t_{c}^{output} - t_{c}^{input}$.

 Lastly, note that for the calculation of carrier frequency shift $D_\omega$ there is no time delay subtraction since this calculation is done entirely in the frequency space where constant time shifts are not relevant.

\section{Systematic and Random Errors in the Experiment}  \label{Errors}
The following sub-sections serve to explain and document the main known possible sources of error in our experiment.
\subsection{Time Domain Equipment limitations}
 The time domain experiment requires coordinated measurements with an Arbitrary Waveform Generator (AWG) and an oscilloscope (DSO). The details of the AWG and DSO models used in the experiments are given in the `Transmission Time Delay Measurements' section of the main text. 

 For the AWG we expect the error in pulse center frequency to be at most $\pm$ 50 kHz ($\pm$ 0.31 $\text{Rad}/\mu s$). This number is estimated based on the age of the internal reference clock of the AWG. The trigger jitter in the DSO is 116 fs (rms).

 To better assess how the limitations of the equipment impacts the experimental data, specifically the center-frequency data, we performed a variety of fidelity measurements. One motivation to examine the accuracy of the center-frequency data was because of the visible noise present in Fig.~4(c) of the main text. The setup for these measurements is depicted in Fig.~\ref{time_delay_subtraction}(a) where we have a single cable connecting the AWG and DSO. A pulse is created in the AWG and sent to the DSO, which performs a triggered measurement of the pulse.  The DSO then performs an averaged measurement over 16 realizations. The final average is considered a single measurement. This process is then repeated 1000 times for various choices of pulse bandwidths. We then take this time domain data, Fourier Transform it in Matlab, and measure the center frequency of the transmitted pulse. The center frequency is calculated using Eqn.~5 of the main text. This value is then compared with the specified center frequency of the original input pulse. The difference between the input and output center-frequencies will be referred to as $D_\omega = \omega_c^{\text{output}}-\omega_c^{{\text{input}}}$. In this scenario, where the pulse is traveling through a single simple coaxial cable, we expect $|D_\omega| \approx 0$.

 \begin{figure}[ht!]
    \centering
    \includegraphics[width=0.48\textwidth]{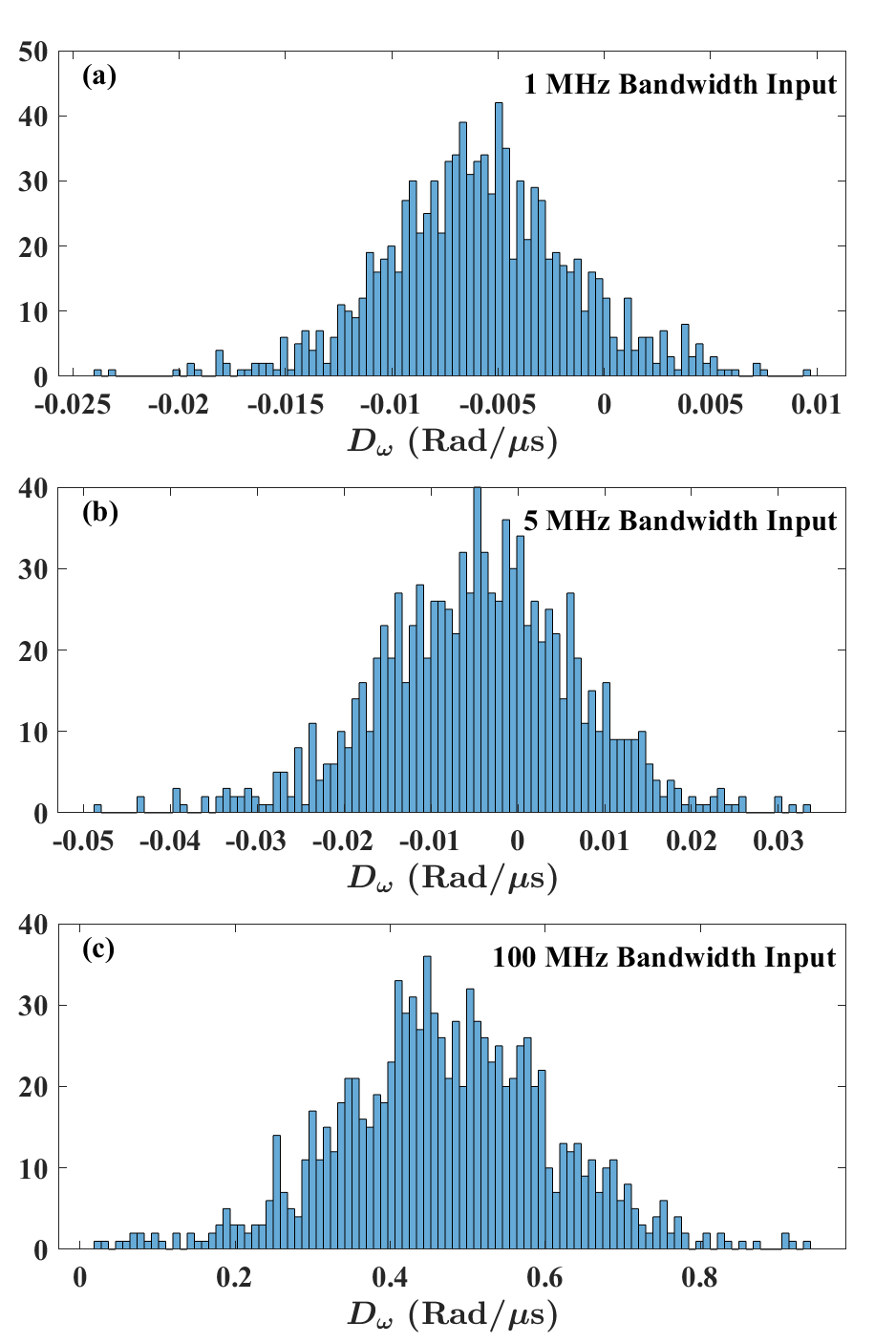}
    \caption{Histograms of $D_\omega$ values measured using the setup in Fig.~\ref{time_delay_subtraction}(a). Each histogram consists of 1000 $D_\omega$ values. The center frequency of the pulse used in each plot is 5.23 GHz. Plots (a-c) are for the cases where the input Gaussian pulse has a bandwidth of 1 MHz, 5 MHz, and 100 MHz respectively.}
    \label{histograms}
\end{figure}

A summary of these results are shown in Fig.~\ref{histograms} for the cases of an input Gaussian pulse with bandwidths of 1 MHz, 5 MHz, and 100 MHz, all centered at 5.23 GHz. For the 1 MHz case, Fig.~\ref{histograms}(a), the mean of the data collected is -0.006 Rad/$\mu$s with a variance of 0.005 Rad/$\mu$s. For the 5 MHz case, Fig.~\ref{histograms}(b), the mean of the histogram is -0.005 Rad/$\mu$s and the variance is 0.011 Rad/$\mu$s. Lastly, for the 100 MHz case, Fig.~\ref{histograms}(c), the mean of the data is 0.472 Rad/$\mu$s and the variance is 0.139 Rad/$\mu$s.  In an ideal scenario, we would expected to have a mean of 0 for all of these cases. In other words, there is no difference between the input Gaussian pulse center frequency and the output Gaussian pulse center frequency when the pulse travels through a single coaxial cable. We see that the means for the 1 MHz and the 5 MHz cases are virtually identical, while for the 100 MHz case the error is significantly larger.

At first glace it might seem that the larger error present in the 100 MHz bandwidth results is problematic, but when looking at the results shown in Fig.~\ref{low_transmission_example}(d) we see that the frequency shifts are on the scale of $10^3$ Rad/$\mu$s. Thus, an error of $\pm$0.472 Rad/$\mu$s is negligibly small in comparison. Similarly, if we look at the results for the 5 MHz bandwidth case shown in Fig.~\ref{low_transmission_example}(c) we see that the frequency shifts are on the order of 1 Rad/$\mu$s. Again this is significantly larger than the measured error of  $\pm$0.005 Rad/$\mu$s. For the 1 MHz bandwidth case, please refer to Fig.~4(c) of the main text. Here we see that the frequency shifts scale on the order of 0.05-0.1 Rad/$\mu$s. The error for this case was $\pm$ 0.006 Rad/$\mu$s. We see in Fig.~4(c) of the main text that the data is noisiest when the magnitude of the measured frequency shifts are roughly less than 0.01 Rad/$\mu$s. Thus, its likely that the ``noise" we see in Fig.~4(c) is simply the result of us reaching the limits of our experiment setup. 

\subsection{Errors arising from differences in time domain and frequency domain experimental setups}
In Fig.~4 (b) and (c) of the main text we see that there is excellent agreement between $D_t$ and Re$[\tau_T]$, and $D_\omega$ and $-\tilde{\Delta}^2\text{Im}[\tau_T]$ when the system is experiencing a strong Feshbach resonance. In between these resonant regions deviations are observed between $D_t$ and Re$[\tau_T]$, and $D_\omega$ and $-\tilde{\Delta}^2\text{Im}[\tau_T]$, where the time domain quantities $D_t$ and $D_\omega$ exhibit oscillatory behavior, but the corresponding Re$[\tau_T]$ and $-\tilde{\Delta}^2\text{Im}[\tau_T]$ quantities derived from calibrated frequency-domain data do not. This behavior has a systematic frequency dependence as shown in Fig.~\ref{full_data}, where at lower frequencies there are less of these oscillations present.

Recall, that the transmission time delay quantities (Re$[\tau_T]$ and $-\tilde{\Delta}^2\text{Im}[\tau_T]$) are calculated directly from the scattering matrix of the system. The scattering matrix of this system is measured in the frequency domain using a network analyzer, see Fig.~1(b) of the main text for a depiction of this measurement setup. It is imperative to note that in this measurement setup the effects of the coaxial cables (that are connecting the ring graph to the network analyzer) are calibrated out using a calibration kit. Thus, the system being measured in the frequency domain is just the ring graph. 

On the other hand, $D_t$ and $D_\omega$ are found by calculating the shifts in transmission time and center frequency of Gaussian pulses as they traverse the graph. These measurements are performed in the time domain using a signal generator and an oscilloscope, see Fig.~\ref{time_delay_subtraction}(c) for an illustration of this setup. For these time domain measurements there is no calibration kit or process that can calibrate out the effects of the external input/output coaxial cables that connects the ring graph to the measurement equipment. Thus, the system being measured is the ring graph plus the input and output cables.


We believe that the discrepancies we see between our frequency domain and time domain derived data in these in-between resonance regions are a result of the fact that the external cabling cannot be calibrated during the time domain measurements.  We were able to confirm these suspicions using both simulation and experiment. 

\begin{figure}[ht!]
    \centering
    \includegraphics[width=0.48\textwidth]{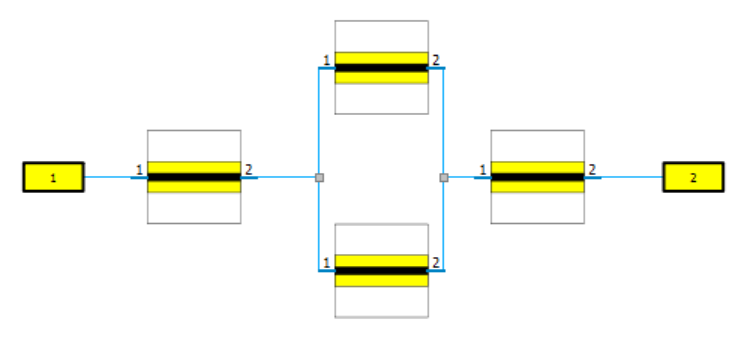}
    \caption{CST simulation model for the case where external cabling is taken into account. Note that the ring graph is identical to the one shown in Fig.~\ref{CST_model_1}.}
    \label{CST_model_2}
\end{figure}

The simulation was modified to include the external cables, as shown in Fig.~\ref{CST_model_2}. The model consists of two ports, the ring graph, and the two external coaxial cables. Note that the ring graph and ports are identical to those utilized in Fig.~\ref{CST_model_1}. The two external cables attached to either side of the ring graph are simulated using the following parameters. They each have a length of 24 inches (61 cm). This is the length of the external cabling used in the experiments. The diameter of the inner and outer conductors are set to 0.0691 cm and 0.205 cm respectively. The relative permittivity of the medium is set to 1.7. The dielectric loss tangent of the medium is set to 0.00005. Lastly, the metal resistivity (normalized to gold resistivity) is set to 1.52. These values were found or estimated using data sheets for the cables used in our experiment.

\begin{figure}[ht!]
    \centering
    \includegraphics[width=0.44\textwidth]{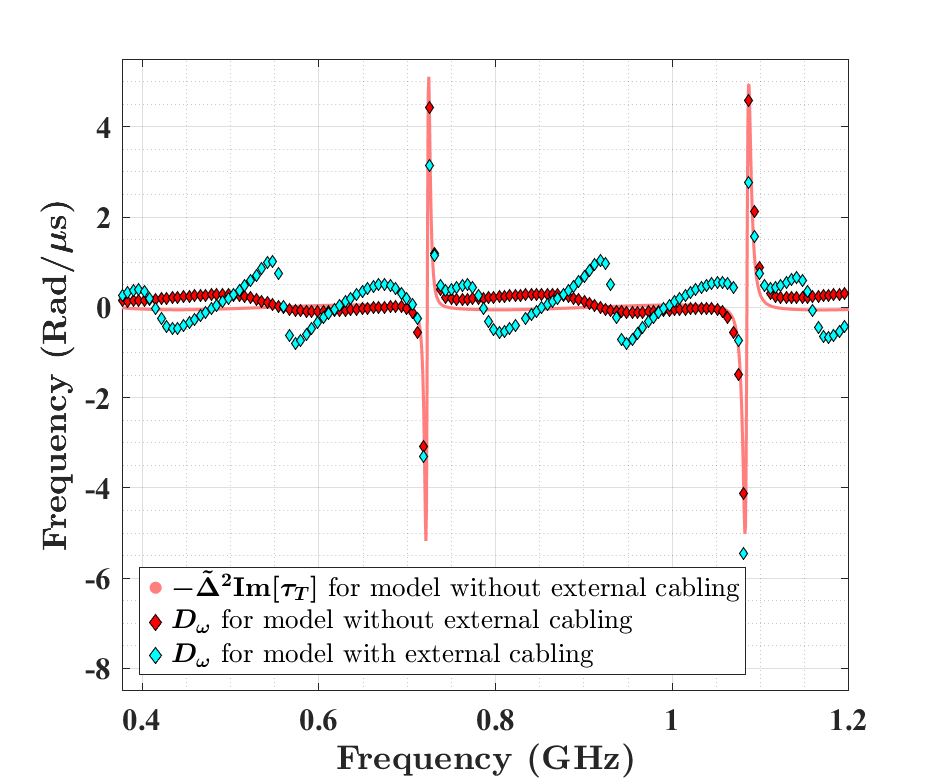}
    \caption{Plot comparing calculated center-frequency shifts for the CST model without external cabling (Fig.~\ref{CST_model_1}) in red with calculated frequency shifts for the CST model with external cabling (Fig.~\ref{CST_model_2}) in cyan. Also shown as a solid light-red line is the predicted center frequency shift $D_{\omega}$ from the CST model without external cabling.}
    \label{ExternalCableResults}
\end{figure}

The main results of this simulation for the center-frequency shift are shown in Fig.~\ref{ExternalCableResults}. The red dots are from Fig.~\ref{simulation_summary}(b), while  the results from the CST simulation with the external cables (Fig.~\ref{CST_model_2}) are plotted on top in cyan. The input Gaussian pulses have a bandwidth of 5 MHz. Here we can clearly see that near the strong resonant frequencies ($\sim$ 0.72 GHz and $\sim$ 1.08 GHz) both simulations follow the $-\tilde{\Delta}^2\text{Im}[\tau_T]$ curve well. In the ``in-between" regions, we see that the cyan curve corresponding to the CST model that takes into account the external cabling has strong oscillations. These oscillations are comparable in character and amplitude to those seen in Fig.~4(d) of the main text. Again note that the CST model that does not take into account the external cabling completely lacks these oscillations.  

Thus we propose that the systematic oscillations seen in the $D_\omega$ data in the experiment arise from standing waves on the input and output cabling.  The observed systematic oscillations generally have a periodicity in frequency of approximately 100 MHz.  Considering a cable of length 24 inches (61 cm), one would expect a half-wave resonance to occur every 174 MHz.  Two such cables will likely produce a pair of such extra resonances, leading to periodicity on the scale of roughly 87 MHz, very much comparable to the data.

\begin{figure}[ht!]
    \centering
    \includegraphics[width=0.55\textwidth]{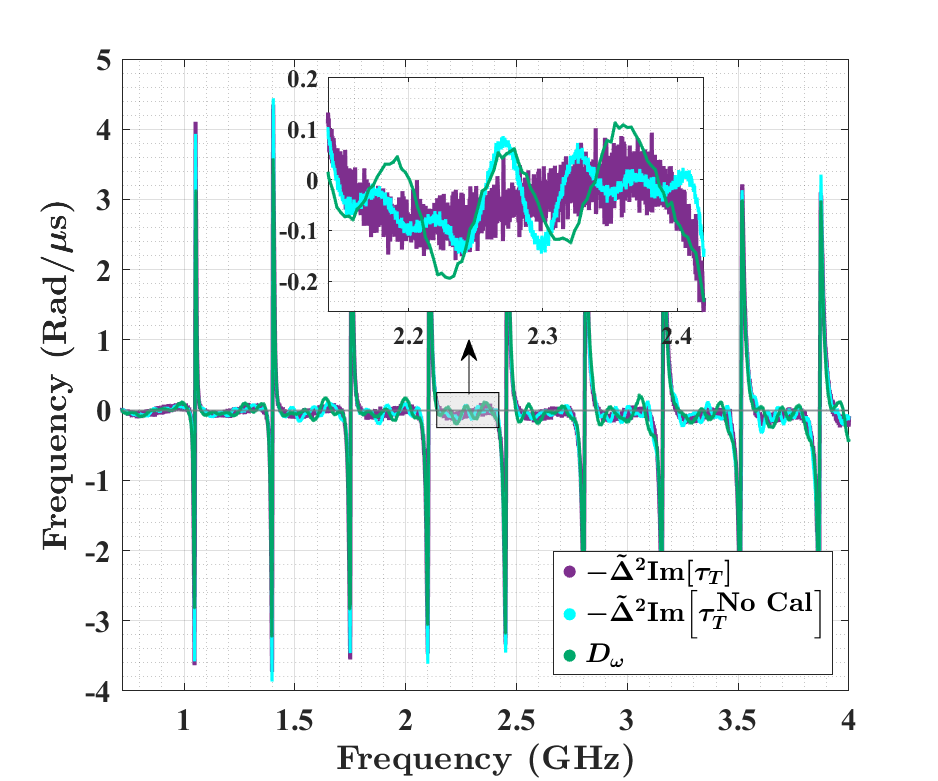}
    \caption{Center-frequency shift data for the case where the input Gaussian pulse has a bandwidth of 5 MHz. In purple, $-\tilde{\Delta}^2\text{Im}[\tau_T]$ is plotted where the external cabling is calibrated out during the frequency domain measurement. In cyan, $-\tilde{\Delta}^2\text{Im}[\tau_T^\text{No Cal}]$ is plotted where the external cabling is not calibrated out. Lastly, in green is the time domain data $D_\omega$ where the external cabling cannot be calibrated out. The inset is a zoomed-in region, of the outer plot, illustrating how $D_\omega$ better follows $-\tilde{\Delta}^2\text{Im}[\tau_T^\text{No Cal}]$.}
    \label{ExternalCableResults_Dw} 
\end{figure}

In addition to these CST simulations, we also performed an uncalibrated measurement of the scattering parameters and used the data $S^{\text{No Cal}}(\omega)$ to recalculate the transmission time delay ($\tau_T^\text{No Cal}$). These results are shown in Fig.~\ref{ExternalCableResults_Dw} and are plotted in cyan. Here we see a clear correspondence between the ``deviations" in the time domain $D_\omega$ and $-\tilde{\Delta}^2\text{Im}[\tau_T^\text{No Cal}]$. This is emphasized by the zoomed-in sub-plot.  This measurement provides further evidence that the systematic oscillations seen in the $D_\omega$ data as a function of pulse center frequency is due to standing waves on the un-calibrated input and output cables.

\section{Further Background Information about Complex Time Delay}  \label{BkdCTD}
Here we are concerned with the presence of loss (or gain) in the system, resulting in a non-Hermitian Hamiltonian and a sub-(super-)unitary scattering matrix. Early theoretical attempts to extend time delay to non-unitary scattering systems related real time delay to the unitary deficit of the S-matrix, \cite{doron1990,Grabsch_2020} and discussed changes to the statistical distribution of the real time delay in over-moded systems \cite{Savin2003}.  However, such systems require a complex generalization of time delay to include the fact that both the phase and the magnitude of the eigenvalues of the $S$-matrix vary with energy.  

 Hints of complex time delay can be found in the use of generalized Wigner-Smith operators to find the `principal modes’ of complicated scattering systems, such as multi-mode optical fibers, \cite{fan2005,xiong2016} particle-like scattering states in multi-mode waveguides, \cite{gerardin2016,Bohm18} and the storage of wave energy in long-lived states in disordered media \cite{Durand19}.  The first explicitly-articulated definition of a complex generalization of time delay for generic non-Hermitian systems appears to be the work of Asano, \textit{et al}., \cite{Asano2016} who consider the complex transmission time delay $\tau_T$ of a simple scattering system.  They consider the propagation of a Gaussian pulse through a waveguide coupled to a ring resonator and show theoretically that $Re[\tau_T]$ describes the time-shift of the center of the pulse, and $Im[\tau_T]$ describes the change in carrier frequency of the pulse after propagating through the system.  They demonstrated experimentally the predicted results for $Re[\tau_T]$ are correct, but did not address the predictions for $Im[\tau_T]$.

 More recently, other researchers have embraced the complex generalization of time delay, finding it useful for description of transmission through disordered media, \cite{Kang2021} where it has been noted that a complex transmission zero leads to a divergence of complex time delay \cite{huang2022,Hougne2021}.  More generally, coherent perfect absorption (CPA), \cite{Chong2010} bringing a complex zero of the S-matrix to the real frequency axis, \cite{Chong2010,Lei2020,Imani2020,Frazier2020,Hougne2021,Erb2024,Erb2024b} has as its signature the divergence of the Wigner-Smith time delay \cite{Li2017,Erb2024}.  Complex time delay can be directly related to the poles and zeros of the non-unitary S-matrix in the complex frequency plane \cite{Lei2021WS,osman2020,Lei2022,Erb2024}.

\newpage
\clearpage
\bibliography{Supplementary}